\documentclass{jaa}
\usepackage{newtxtext,newtxmath}
\usepackage[T1]{fontenc}
\usepackage{ae,aecompl}
\usepackage{graphicx}
\usepackage{caption}
\usepackage{subcaption}
\DeclareUnicodeCharacter{2212}{-}

\begin{document}\sloppy

\title{The Horizontal Branch morphology of the globular cluster NGC 1261 using AstroSat}


\author{Sharmila Rani\textsuperscript{1,2,*}, Gajendra Pandey\textsuperscript{1}, Annapurni Subramaniam\textsuperscript{1}, Snehalala Sahu\textsuperscript{1} and N. Kameswara Rao\textsuperscript{1}}
\affilOne{\textsuperscript{1}Indian Institute of Astrophysics, Koramangala II Block, Bengaluru, 560034, India\\}
\affilTwo{\textsuperscript{2}Pondicherry University, R.V. Nagar, Kalapet, 605014, Puducherry, India}


\twocolumn[{

\maketitle

\corres{sharmila.rani@iiap.res.in}

\msinfo{06 September 2020}{24 September 2020}

\begin{abstract}
We present the results obtained from the UV photometry of the globular cluster NGC 1261 using Far-UV (FUV) and Near-UV (NUV) images acquired with the Ultraviolet Imaging Telescope (UVIT) on board the ASTROSAT satellite. We utilized the UVIT data combined with HST, GAIA, and ground-based optical photometric data to construct the different UV colour-magnitude diagrams (CMDs). We detected blue HB (BHB), and two extreme HB (EHB) stars in FUV, whereas full HB, i.e., red HB (RHB), BHB as well as EHB is detected in NUV CMDs. The 2 EHB stars, identified in both NUV and FUV, are confirmed members of the cluster. The HB stars form a tight sequence in UV-optical CMDs, which is almost aligned with Padova isochrones. This study sheds light on the significance of UV imaging to probe the HB morphology in GCs.
\end{abstract}


}]


\doinum{12.3456/s78910-011-012-3}
\artcitid{\#\#\#\#}
\volnum{000}
\year{0000}
\pgrange{1--}
\setcounter{page}{1}
\lp{1}

\section{Introduction}
Globular clusters (GCs) are the ideal laboratories to study the formation and evolution of different stellar populations
as they are the oldest objects known to exist in our galaxy.
They provide the best platform to study exotic populations such as blue straggler stars (BSSs), cataclysmic variables, low mass X-ray binaries (LMXB), etc. NGC 1261 is an old, metal-intermediate ([Fe/H] = -1.42 dex) GC located in the constellation Horologium at a distance of 17.2 kpc (Arellano Ferro {\em et al.} 2019).\\

Optical imaging using the  Hubble Space Telescope has been carried out by Piotto {\em et al.} (2002), along with other 74 Galactic GCs. They constructed the optical colour-magnitude diagrams (CMDs) using HST F439W and F555W bandpasses. Kravtsov {\em et al.} 2010 studied the wide-field multi-colour photometry of this cluster using data from 1.3 m Warsaw telescope at Las Campanas Observatory. They mainly focused on the brighter sequences of the cluster’s CMD. Identification of hot exotic populations in optical images is difficult as optical images are more crowded than UV images, dominated by main-sequence (MS) and red giants (RGBs), and the hot stars are optically faint. UV study of evolved populations in GCs is very important because a few critical evolutionary phases, found in the GCs, are brighter in UV than in optical wavelength. The main contributors to the luminosity of GCs in the UV are horizontal branch (HB) stars, BSSs, post-asymptotic giant branch (PAGB) stars, and white dwarfs (WDs) (Zinn {\em et al.} 1972; Harris {\em et al.} 1983). UV colour-magnitude diagrams (CMDs) play a vital role to identify and study the properties of UV bright stellar populations in GCs ( Ferraro {\em et al.} 2003; Haurberg {\em et al.} 2010; Dieball {\em et al.} 2010).
The HB stars are core helium burning low mass stars with a core mass
of approximately $\sim0.5M_{\odot}$ (Iben \& Rood 1970). The different parts of the HB are red HB (RHB), blue HB (BHB), and extreme HB (EHB). RR Lyrae instability strip separates the RHB stars from the BHB stars. EHB stars are core helium-burning stars with an envelope too thin to sustain hydrogen burning. These stars are expected to lie at the blue/hot end of the blue tail of the HB in the optical CMDs, and their effective temperature is found to be more than 20,000 K. The UV wavelength acts as an excellent probe to study the HB stars and their characteristics, as these star appear brighter in the UV.\\ 

\begin{figure}
\centering
\includegraphics[scale=0.42]{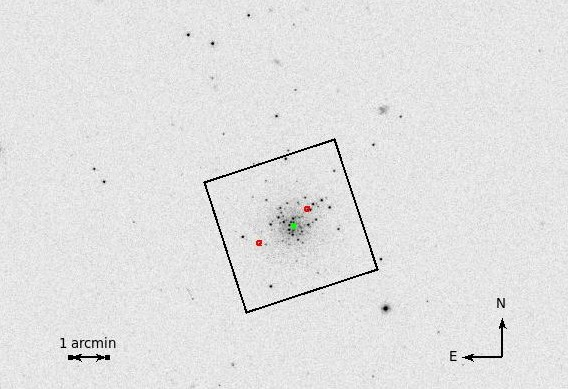}
     \caption{UVIT FUV (F172M) image of NGC 1261. The green cross symbol corresponds to the center of the cluster, and the black square box represents the area covered with HST (3.4$'$). The locations of two EHB stars detected with UVIT are shown with red open circles.}
     \label{combimage}
\end{figure}

\begin{table*}
	\centering
	\caption{Details of basic parameters and observations of NGC 1261}
	\label{tab:1}
	\begin{tabular}{cccccccc} 
		\hline
		\hline
 Cluster ID & R.A. (J2000)& Decl. (J2000) &  \textit{l} & \textit{b} & Filter & $\lambda_{mean}$ & Exposure\\
 & (deg) & (deg)  & (deg) & (deg) & & ({\AA}) & Time (sec)\\
		\hline 
		 NGC 1261 & 48.0675 & $-$55.2162 & 270.5387 & −52.1243 & F169M & 1608 & 1746\\
		 & & & & & F172M & 1717 & 6662\\
		 & & & & & N219M & 2196 & 2847\\
		 & & & & & N245M & 2447 & 740\\
		 & & & & & N263M & 2632 & 1022\\
		 & & & & & N279N & 2792 & 3831\\
		\hline
	\end{tabular}
\end{table*}
 Brown {\em et al.} 2016 studied the HB morphology in 53 GCs, including NGC 1261. They found less than four HB stars blue-ward of the G-jump in NGC 1261 (see Figure 5 in Brown {\em et al.} 2016). Schiavon {\em et al.} 2012 constructed UV CMDs (FUV$-$NUV vs. FUV) for 44 galactic globular clusters using GALEX data. They found that the HB stars follow a diagonal sequence, unlike the horizontal distribution in optical and  its slope mainly depends on the bolometric correction effects.\\


The BSSs are defined as the stars brighter and bluer than the main-sequence (MS) turn-off in the colour-magnitude diagram of the star clusters. The standard theory of stellar evolution does not explain the origin of BSSs. The main leading scenario suggested for the BSS formation in star clusters are (1) mass transfer in close binary systems ( McCrea {\em et al.} 1964) and (2) merger of binaries induced by collisions ( Hills \& Day {\em et al.} 1976). In star clusters, the formation scenario due to stellar collisions dominates in high-density environments, whereas other formation scenarios dominate in low-density environments. UV photometry helps to separate BSSs from the hot HB stars, as the hot HB stars are very much luminous in the UV, when compared to the relatively cooler BSSs.

In this study, we present the results of a UV imaging study of the GC NGC 1261 in six filters (2 FUV and 4 NUV), using the Ultra-violet Imaging Telescope (UVIT) on ASTROSAT. We used the proper motion estimated using HST and GAIA data to select cluster members in the inner and outer region of the cluster, respectively. We detect the
HB morphology, including a few UV bright stars among the HB population.

\section{Observations and Data Analysis}
The data analyzed in this study are taken from UVIT onboard AstroSat, the first Indian space observatory. Observations of the cluster, NGC 1261, were performed on 26 August 2017 by UVIT in two far-UV (FUV) and four near-UV (NUV) filters. The spatial resolution of UVIT is better than $1.8"$ in both FUV and NUV bands. We obtained the magnitudes in all filters by performing a psf photometry on the science ready UV images using the DAOphot package in IRAF. The aperture and saturation corrections were also done to obtain the final magnitudes in each filter (Tandon {\em et al.} 2017b). The instrumental magnitudes were calibrated to the AB magnitude system using zero-points reported in Tandon {\em et al.} 2017b. The stars as faint as 21 magnitude in NUV and 22 magnitude in FUV with typical errors 0.2 and 0.3 mag are detected.The details of the basic parameters and UVIT observations of the cluster are tabulated in Table~\ref{tab:1}. The FUV image of the cluster in the F172M filter of the UVIT is shown in Figure~\ref{combimage}. In  this image, we show the area of the UVIT image used in this study, and the region covered with HST with black square box.

\begin{figure*}
\centering
\begin{subfigure}{0.5\textwidth}
  \centering
  \includegraphics[height=8.0cm,width=9.0cm]{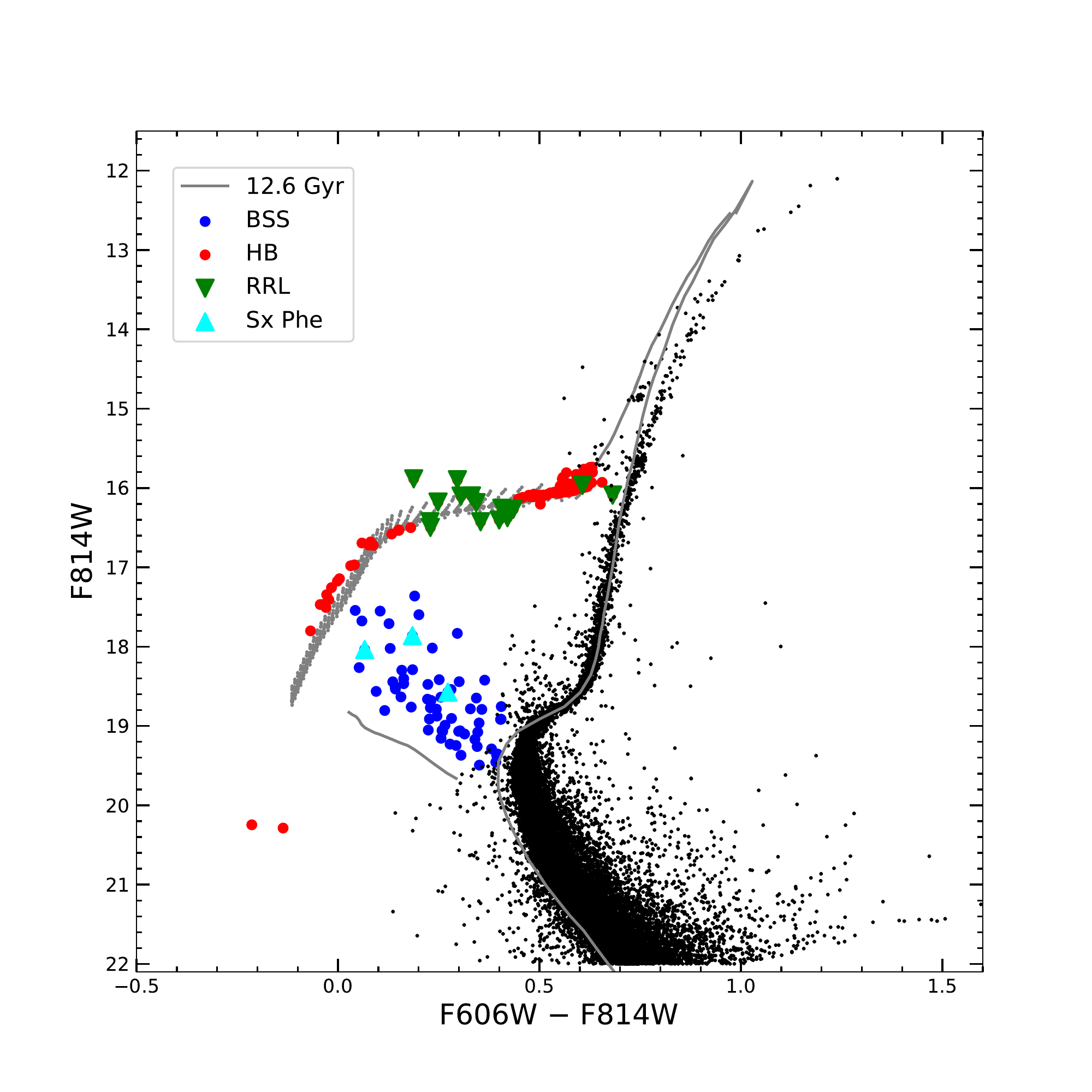}
  \end{subfigure}%
\begin{subfigure}{0.5\textwidth}
  \centering
  \includegraphics[height=8.0cm,width=9.0cm]{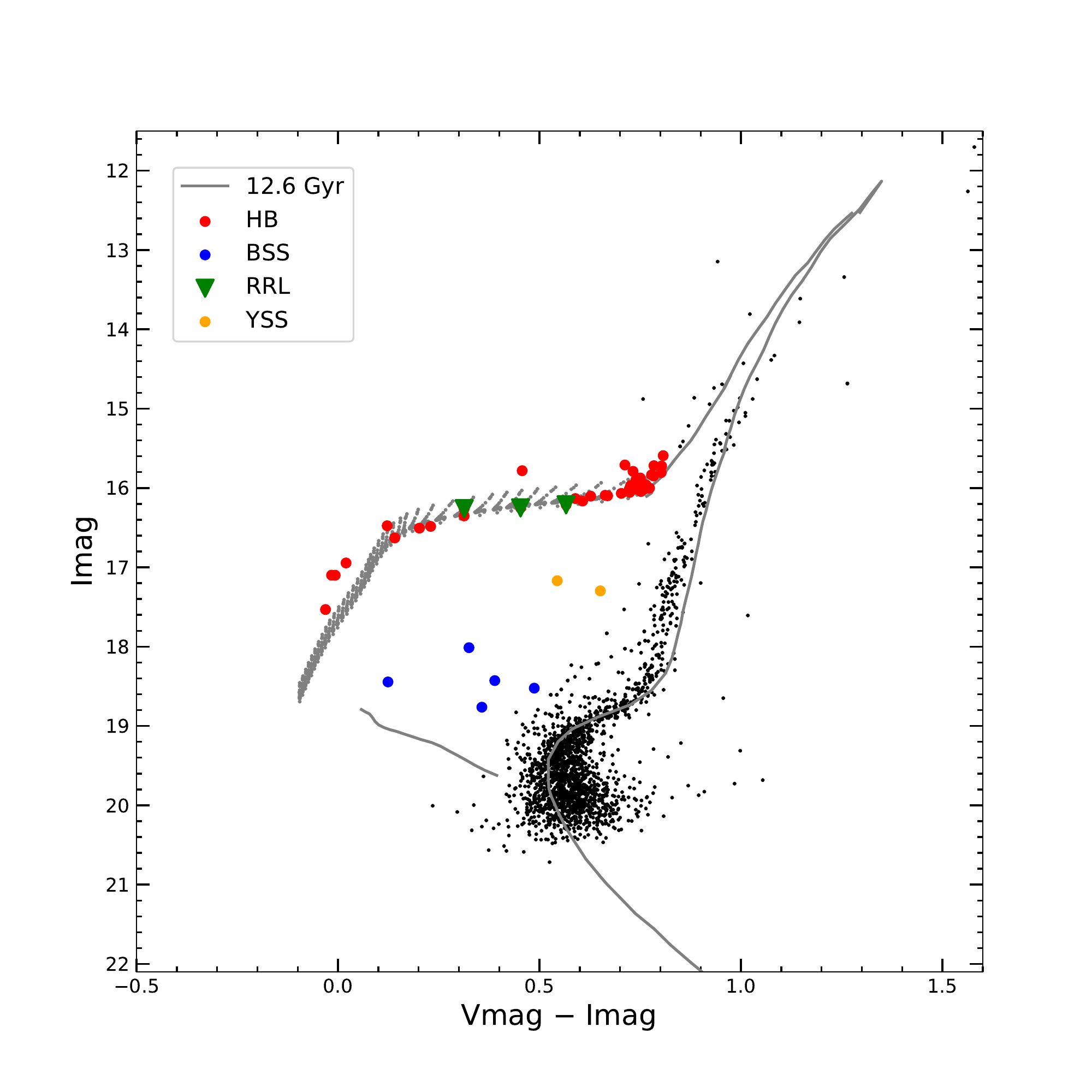}
\end{subfigure}
\caption{(Left panel) Optical CMD of NGC 1261 for the inner region. Black dots correspond to the proper motion (PM) members detected with HST. Only HB and BSS stars detected with both UVIT NUV N279N and HST are shown with red and blue filled circles. The known variable stars, such as RR Lyrae (green down triangles) and SX Phe (cyan up triangles), are also shown in the figure. The solid grey lines denote Padova model isochrone for 12.6 Gyr and \big[Fe/H\big]= -1.42 dex. The BSS model line shown in the figure is an extension of the zero-age main sequence (ZAMS). (Right panel) Optical CMD of NGC 1261 for the outer region. Only HB and BSS stars detected with UVIT NUV N279N, ground, and Gaia are shown. The rest of the stars shown with black dots are cross-matched ground-based data with Gaia data. The rest of the details are the same as in left panel.}
\label{optcmds}
\end{figure*}

\begin{figure*}
   \hspace*{-1.4cm} 
\includegraphics[height=20cm,width=20cm]{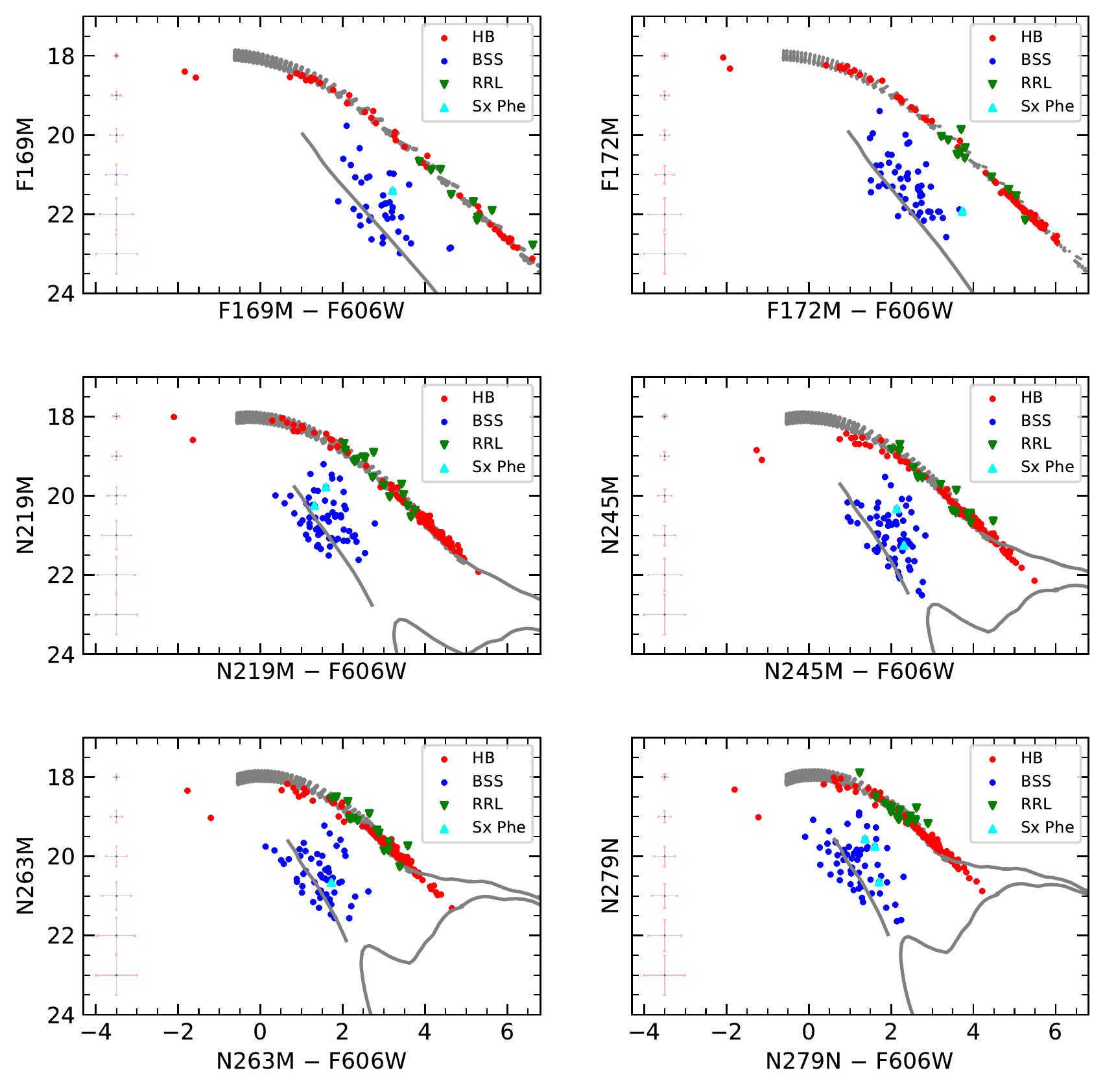}
     \caption{UV-optical CMDs of NGC 1261 after cross-matching HST data with UVIT data in 4 NUV and 2 FUV filters. The meaning of different colours and symbols is shown in the panels. The photometric errors in magnitude and colour are also shown in each panel. The solid grey lines denote the Padova model isochrone for 12.6 Gyr and \big[Fe/H\big] = $-$1.42 dex.}
    \label{optuvhstcmds}
\end{figure*}

\begin{figure*}
\hspace*{-1.4cm} 
\includegraphics[height=20cm,width=20cm]{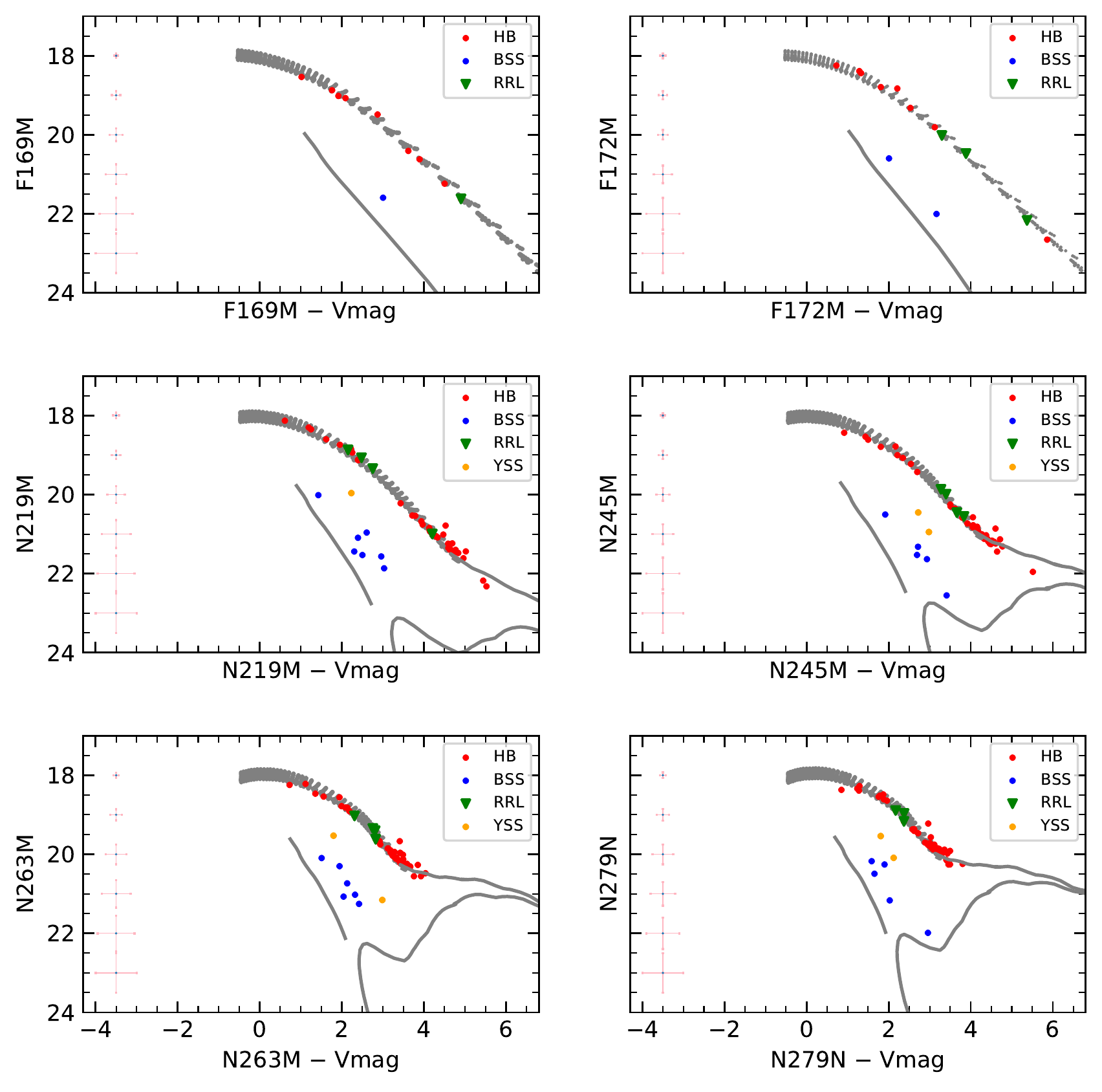}
     \caption{UV-optical CMDs of NGC 1261 after cross-matching ground based photometric data and Gaia data with UVIT data in 4 NUV and 2 FUV filters. Rest of the details are same as in Figure~\ref{optuvhstcmds}.}
    \label{optuvgaiacmds}
\end{figure*}

\subsection{UV and Optical Colour-Magnitude Diagrams}
To identify different stellar populations detected with UVIT in the inner region (within 3.4$'$ diameter), we cross-matched UVIT detected stars with HST UV legacy survey catalog of GCs (Nardiello {\em et al.} 2018). The proper motion (PM) membership probability for the HST detected stars is provided by Nardiello {\em et al.} 2018, and they mentioned that the most likely members of the cluster have PM membership probability greater than $90\%$. The common stars detected with both HST and UVIT are identified, and the proper motion (PM) members of the cluster are selected for further study. To adopt the same magnitude system, we transformed the VEGA magnitude system (of the HST) into the AB magnitude system. We cross-matched the variable star catalog provided by Bustos Fierro {\em et al.} 2019 with UVIT data to identify the variable stars including RR Lyrae and SX Phe. Our sample of HB stars is not complete in the inner region as we are not able to resolve inner $1'$ region due to crowding and outside the $1'$ region, we detected about 90\% stars compared to the HST. We constructed the PM cleaned optical and UV-optical CMDs for the members in the inner region 
as shown in (left panel) Figure~\ref{optcmds} and Figure~\ref{optuvhstcmds}, respectively. The over-plotted solid grey lines correspond to a Padova model isochrone for the metallicity [Fe/H] = -1.42 dex and age 12.6 Gyr generated using flexible stellar population synthesis (FSPS) code (Conroy {\em et al.} 2009). The grey solid line shown along BS locus is the extrapolation of the zero-age main-sequence (ZAMS). It is observed from the UV-optical CMDs that the HB stars are found to be located along a diagonal sequence, instead of the horizontal sequence as found in optical CMDs, 
(Subramaniam {\em et al.} 2017; Sahu {\em et al.} 2019). We detected two stars at the blue end of the HB in both FUV and NUV CMDs (See figure~\ref{optuvhstcmds}). From the UV-optical colour of these two stars, we suggest that these stars are likely to be the hotter counter-parts of the HB. Only BHB stars are detected in FUV as these stars are hot enough to emit at this wavelength. We detected full stretch of the HB including RHB, RR Lyrae and BHB in NUV-optical CMDs shown in lower four panels in Figure~\ref{optuvhstcmds}. We observe that the HB stars form a tight sequence well aligned with Padova isochrones in two FUV-optical and one NUV-optical CMD (see middle left panel). In rest of the NUV-optical CMDs, the observed HB sequence is more-or-less aligned with isochrone, except the red end of the HB which is fainter than the isochrone. This might be due to the large photometric errors at this magnitude. The HB stars and BSSs found to span a wide range in colour as well as in magnitude in all UV-optical CMDs. The detected BSSs are found to be as blue as the BHB stars in NUV-optical CMDs.\\

In the case of outer part of the cluster (outside the 3.4$'$ diameter), optical photometric data obtained from the ground (Kravtsov {\em et al.} 2010) is cross-matched with UVIT data. In order to find the PM membership of the detected stars in the outer region, we then cross-matched with Gaia PM membership data provided by Bustos Fierro {\em et al.} 2019. Our sample of HB and BSSs detected with UVIT is 90\% complete in the outer region, when compared to the number of stars detected with Gaia. The 10\% of the stars, which are not detected, are fainter than the detection limit of the UVIT. The optical and UV-optical CMDs for the outer region are shown in (right panel) Figure~\ref{optcmds} and Figure~\ref{optuvgaiacmds}, respectively. 
We also notice that, in the upper two FUV panels, only a few BHB stars are detected and in lower four NUV panels, both RHB and BHB stars are detected. 
Isochrones are almost aligned with observed HB pattern in all the UV-optical CMDs. Candidate Yellow straggler stars (YSSs) are also detected in all NUV-optical CMDs, identified from their location in optical CMDs. From the comparison of the BSSs detected in inner and outer region, we suggest that BSSs may be segregated towards the center of the cluster. This requires a detailed analysis and will be performed in the future.\\

We have detected two EHB stars both in the NUV and the FUV images. These stars are located well away from the BHB stars and are members of the cluster.  The location of these two stars is shown in the UVIT image in Figure~\ref{combimage}. It will be important to understand the properties of these stars, such as their Luminosity, temperature, radius etc, to identify their evolutionary nature. Placing them on the H-R diagram along with the evolutionary models will help in estimating their mass. This is likely to place a limit on the amount of mass loss it must have experienced in the RGB phase. The chemical composition of these EHB stars  using spectroscopy needs to be performed in order to constrain their surface chemical composition.
From the chemical abundances and effective temperatures of the stars, we can determine their evolutionary status, for example, whether the stars are still in the HB phase or they are evolving into to the post-EHB phase. As this evolutionary phase is short lived and hence less understood, we believe that the identification of EHB stars in the cluster is the first step in this direction.  This study therefore points to the usefulness of UV images to study the HB morphology of GCs and to detect hot HB stars.\\ 

The formation pathway of EHB stars continue to be debated and to address this, a complete census of EHB stars in GCs is needed. In the HB studies of GCs using the AstroSat/UVIT, Sahu et al. (2019) found 3 such stars in the moderately dense NGC 288. The less dense cluster, NGC 5466, appears to be devoid of EHB stars (Sahu \& Subramaniam 2020). In the relatively dense cluster, NGC 1851, with its large number of HB stars spanning a large colour range in UV  (Subramaniam et al. 2017), a few EHB stars are detected and are studied in detail by Singh et al. 2020, (submitted).\\ 

Multiple scenarios have been suggested to explain the formation of the EHB stars in GCs. Still, not a single formation scenario is able to reproduce all the observed characteristics of these stars. The main challenge is to explain the large mass loss required for the formation of EHB stars. There are different theories explaining EHB stars through various processes at the end or after the RGB. The different formation scenarios suggested for the EHB stars in GCs are helium mixing, late hot-flasher, helium-enrichment, mass loss through rapid rotation, white-dwarf mergers, etc. (D’Cruz {\em et al.} 1996; Sweigart {\em et al.} 1997; Saio \& Jeffery 2000; Brown {\em et al.} 2001, 2010). Our ongoing study of GCs using ASTROSAT is expected to increase the number of known EHB stars to throw light on their formation pathways.

\vspace{1em}
\section{Conclusion}
We investigate the HB morphology of the GC NGC 1261 imaged using the UVIT on AstroSat observatory, and further created the optical and UV-optical CMDs of the PM member stars co-detected using UVIT and HST, in this cluster. For comparison with theoretical predictions, we overlaid the CMDs with Padova model isochrones generated for respective UVIT and HST filters. Only hot HB stars are detected in FUV as the cooler RHB stars are too faint to be detected in FUV. We detect two UV bright stars, located at the blue end of the HB. These two EHB stars detected with UVIT are ideal candidates to further study their properties including chemical abundances. This study thus demonstrates the importance of using UV photometry to examine the HB morphology and to detect hot HB stars.

\vspace{1em}

\section*{Acknowledgements}
UVIT project is a result of collaboration between IIA, Bengaluru,  IUCAA,  Pune,  TIFR,  Mumbai,  several  centres  of ISRO, and CSA. This publication uses the data from the \textit{ASTROSAT} mission of the Indian Space Research  Organisation  (ISRO), archived at the Indian Space Science Data Center (ISSDC).

\vspace{-1em}



\end{document}